\documentclass[preprint,showpacs,preprintnumbers,amsmath,amssymb,superscriptaddress,jap]{revtex4}

\usepackage{graphicx,amsmath,amsfonts,amssymb,fancybox}
\usepackage{dcolumn}
\usepackage{bm}
\usepackage{color}
\usepackage[usenames,dvipsnames]{xcolor}

\begin{document}

\title{Density Functional Theory Calculation of Edge Stresses in Monolayer MoS$_{2}$}

\author{Zenan Qi\footnote{These authors contributed equally to this manuscript.}}
    \affiliation{Department of Mechanical Engineering, Boston University, Boston, MA 02215}
\author{Penghui Cao\footnote{These authors contributed equally to this manuscript.}}
    \affiliation{Department of Mechanical Engineering, Boston University, Boston, MA 02215}    
\author{Harold S. Park\footnote{Electronic address: parkhs@bu.edu}}
    \affiliation{Department of Mechanical Engineering, Boston University, Boston, MA 02215}

\date{\today}

\begin{abstract}

We utilize density functional theory to calculate the edge energy and edge stress for monolayer MoS$_{2}$ nanoribbons.  In contrast to previous reports for graphene, for both armchair and zigzag chiralities, the edge stresses for MoS$_{2}$ nanoribbons are found to be tensile, indicating that their lowest energy configuration is one of compression in which Mo-S bond lengths are shorter than those in a bulk, periodic MoS$_{2}$ monolayer.  The edge energy and edge stress is found to converge for both chiralities for nanoribbon widths larger than about 1 nm. 

\end{abstract}


\maketitle

Two-dimensional materials such as graphene have been extensively studied in recent years, owing to their exceptional mechanical~\cite{leeSCIENCE2008}, electrical~\cite{netoRMP2009}, and other physical properties.  However, because graphene is gapless, researchers investigated the electronic properties of graphene nanoribbons, which become semiconducting as the nanoribbon width becomes sufficiently small~\cite{hanPRL2007,liSCIENCE2008,wangPRL2008}.  Due to the introduction of edges and dangling bonds in nanoribbons, edge effects on the mechanical~\cite{huangPRL2009,junPRB2008,shenoyPRL2008} and electrical properties of graphene nanoribbons  have subsequently been widely studied~\cite{hanPRL2007,liSCIENCE2008,wangPRL2008}.  

Also motivated by graphene's gaplessness, research on two-dimensional crystals has recently turned to the transition metal dichalcogenides (TMDs), and specifically MoS$_{2}$, which was recently found to have a direct gap of nearly 2 eV in two-dimensional monolayer form~\cite{makPRL2010}.  Since then, there have been many studies on using MoS$_{2}$ for nanoelectronics~\cite{radNNANO2011}, and other applications~\cite{atacaJPCC2012,chhowallaNC2013,wangNNANO2012}.  It has also been found through theoretical calculations that MoS$_{2}$ nanoribbons exhibit interesting electronic properties~\cite{atacaJPCC2011,liJACS2008}, specifically ferromagnetic and metallic behavior~\cite{liJACS2008}.  

While the properties of MoS$_{2}$ nanoribbons are of interest, very few studies on the edge elastic and mechanical properties of MoS$_{2}$ exist, with the exception of a recent work~\cite{dengAPL2012} that focused on the edge stresses of non-stoichiometric edges, and another that calculated the size-dependent Young's modulus of MoS$_{2}$ nanoribbons~\cite{jiangJAP2013}.  The mechanical characterization of the edges of MoS$_{2}$ is also important due to the interest in tailoring the electronic properties of MoS$_{2}$ via mechanical strains~\cite{johariACSNANO2012}.  Therefore, the purpose of this work is to, using density functional theory (DFT) calculations, characterize the edge energies and stresses for the armchair and zigzag directions of monolayer MoS$_{2}$.


To compute the total energy, edge energy and edge stress, we conducted DFT calculations using the open-source code SIESTA~\cite{solerJPCM2002}.  In detail, we used the norm-conserving nonlocal Troullier-Martins pseudopotential~\cite{troullierPRB1991} and the local density approximation (LDA) parameterized by~\citet{perdewPRB1981}. Double-$\zeta$ polarized basis sets were used for the valence electrons of both Mo and S with an energy shift parameter of 0.01 Ry~\cite{artachoPSSB1999}, while the energy cutoff was set to be 250 Ry for real-space integration.  To determine the energy of the MoS$_{2}$ nanoribbons, we used a Monkhorst-Pack scheme and a $2\times2\times2$ $k$-point mesh to assure convergence and for computational efficiency; we verified that using more $k$-points ($5\times5\times2$ and $10\times10\times2$) did not change our results.  The density matrix tolerance was set to be 5$\times 10^{-5}$ and the maximum number of iterations was chosen as 300. The density matrix mixing weight parameter was set as 0.01 and the Pulay number was set as 5 to accelerate the convergence.  The diagonalization method was used to solve the Kohn-Sham equations and the relaxed atomic positions were found using the conjugate gradient method until the forces on each atom were smaller than 0.02 eV/\AA. After relaxation, we obtained the lattice constant of monolayer MoS$_{2}$ to be 3.125~\AA~with a monolayer thickness of 3.21~\AA, which is consistent with previous DFT and experimental results~\cite{bokerPRB2001,cooperPRB2013,atacaJPCC2011,liJACS2008,molinaPRB2011}.  

\begin{figure} \begin{center} 
\includegraphics[scale=0.38]{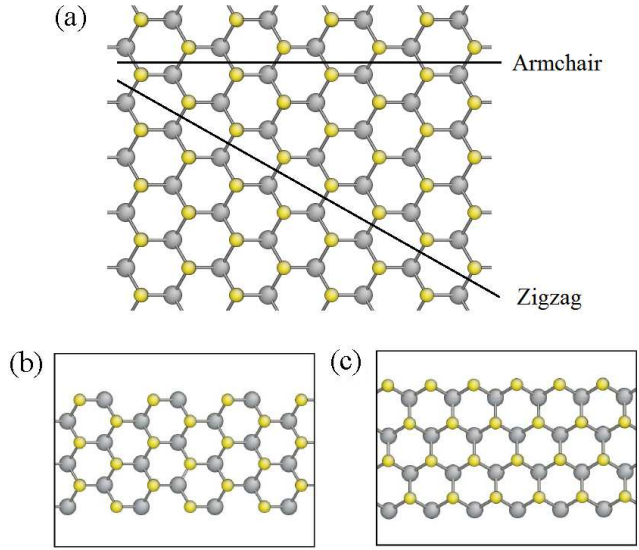}
\caption{(Color online) (a) Illustration of how zigzag and armchair MoS$_{2}$ nanoribbons are formed from bulk monolayer MoS$_{2}$.  (b) Schematic of armchair MoS$_{2}$ nanoribbon (AMSNRs) before relaxation. (c)  Schematic of zigzag MoS$_{2}$ nanoribbon (ZMSNRs) before relaxation.  Gray and yellow atoms represent Mo and S, respectively, and the white space above and below the nanoribbons represents free space to avoid interaction with other periodic super cells.}
\label{fig1} \end{center} \end{figure}

The edge stress $f$ is defined following the work of~\citet{junPRB2008}
\begin{equation}\label{eq:edge1} \gamma(\epsilon)=\gamma_{0}+f\epsilon,
\end{equation}
where $\gamma$ is the edge energy, $\gamma_{0}$ is the edge energy when the edge is unstrained, and $\epsilon$ is the strain.  Physically, the edge energy $\gamma$ is the energy required to form a new edge, while the edge stress $f$ represents the unit work required to deform the existing edge.
The edge energy $\gamma$ also represents the total excess energy, or difference in energy between atoms at the edge as compared to atoms within the bulk, per edge length $L$.  The excess edge energy and stress both originate from the fact that edge atoms have a lower coordination number, or fewer bonds than atoms within the MoS$_{2}$ bulk.  

To calculate the edge energy $\gamma$, we define
\begin{equation}\label{eq:edge2} \gamma=\frac{1}{2L}(E_{ribbon}^{N}-Ne_{pbc}),
\end{equation}
where $E_{ribbon}^{N}$ is the total energy of the MoS$_{2}$ nanoribbon, $L$ is the edge length, $e_{pbc}$ is the total energy per atom of the periodic MoS$_{2}$ monolayer, and $N$ is the number of atoms in the nanoribbon.  The edge stress $f$ can be calculated as $f =\left(\gamma(\epsilon) - \gamma_0\right)/\epsilon$. 

Eq. (\ref{eq:edge2}) states that, in our simulations, the edge energy $\gamma$ is calculated from the total energy difference between monolayer MoS$_{2}$ with and without edges.  Therefore, flat MoS$_2$ armchair and zigzag nanoribbons of varying widths were generated by first cutting from a periodic MoS$_2$ monolayer, as shown in Fig. \ref{fig1}(a).  These nanoribbons were then placed into three-dimensional unit cells with thickness and edge directions separated by more than 10~\AA~to avoid spurious interaction between super cells as shown in Fig.\ref{fig1}. The relaxed nanoribbon configurations were obtained by minimizing their total energies.  The tensile and compressive strain needed to evaluate the edge stress in Eq. (\ref{eq:edge1}) was then applied to the relaxed nanoribbon.

As shown in Fig. \ref{fig1}(b), armchair MoS$_{2}$ nanoribbons (AMSNRs) have alternating Mo and S atoms on both edges, while zigzag MoS$_{2}$ nanoribbons (ZMSNRs) as shown in Fig. \ref{fig1}(c) have all Mo atoms on one edge with all S atoms on the other edge, which is the most stable and energetically favorable ZMSNR~\cite{dengAPL2012}.  Because the nanoribbons have different atomic surface terminations, the edge energy $\gamma$ and edge stress $f$ we report represent averages of the two different surfaces, similar to what has been done previously for multi-atom nano structures such as boron nitride (BN)~\cite{junPRB2011}.  Without external strains, we found the relaxed edge Mo-S bond length to be 2.280~\AA~for AMSNRs of width 9.37~\AA~and 2.370~\AA~for ZMSNRs of width 9.02~\AA, which are both shorter than the bulk monolayer Mo-S bond length of  2.393~\AA.  

\begin{figure} \begin{center} 
\includegraphics[scale=0.8]{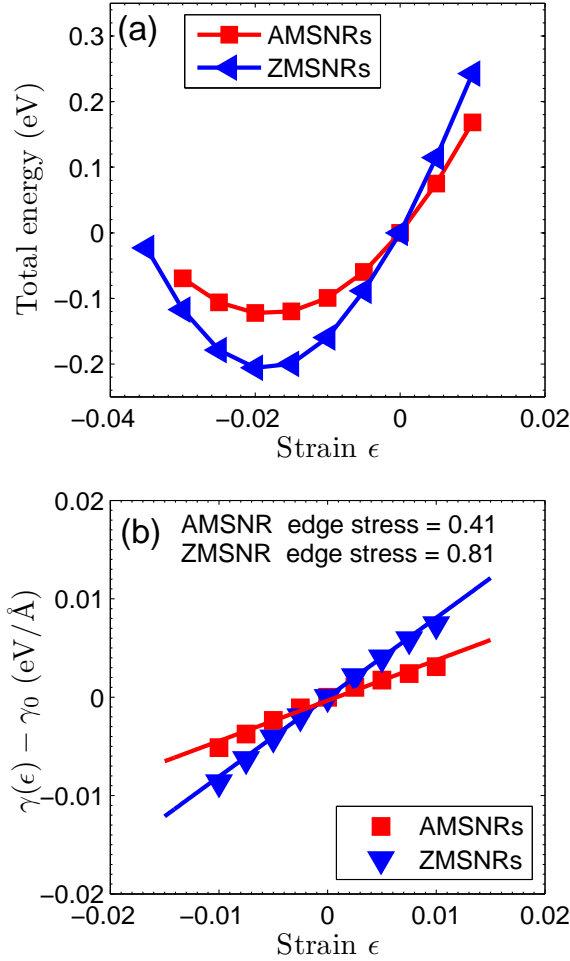}
\caption{(Color online)(a) Total energy of ZMSNR of width 9.02~\AA~and AMSNR of width 9.37~\AA~versus strain. (b)  Relative edge energy versus strain, where the slope of the line represents the edge stress $f$.}  
\label{fig2} \end{center} \end{figure}

We begin our discussion of the results by showing in Fig. \ref{fig2} the total energy and relative edge energy $(\gamma(\epsilon)-\gamma_{0})$ as functions of applied uniaxial strain for an AMSNR of width 9.37~\AA, and a ZMSNR of width 9.02~\AA.  Fig. \ref{fig2}(a) shows the total energies of both the AMSNR and ZMSNR as a function of strain.  Of note, the minimum energy for both chiralities occurs at a negative (compressive) strain, which happens to be about -2\% for both nanoribbons for the widths of 9.02~\AA~and 9.37~\AA~ for ZMSNR and AMSNR, respectively.  

To calculate the edge stress, uniaxial tensile and compressive strain increments of 0.25\% were applied to the nanoribbons up to a total strain of -1\% to 1\%.

The edge energy for each state of strain was calculated and fit to a linear function to compute the edge stress $f$ as discussed following Eq. \ref{eq:edge1} above.  This result is consistent with the result shown in Fig. \ref{fig2}(b), in which the edge stress for the two nanoribbon chiralities and widths is found to be positive, i.e. 0.81 eV/\AA~for the ZMSNR and 0.41eV/\AA~for the AMSNR; these values are slightly larger than previous study by~\citet{dengAPL2012} as they considered triangular edges instead of ideal ones. We note that the edge stress fitting begins to deviate from linearity for the larger tensile and compressive strains in Fig. \ref{fig2}(b), where similar effects have been previously reported for graphene~\cite{luMSMSE2011}.

The positive value of the edge stress means that the edges can minimize their energy by making their bond lengths smaller, which leads to the minimum energy being at a compressive strain as shown in Fig. \ref{fig2}(a).  Interestingly, the sign of the edge stress is opposite to that of graphene, for which a negative edge stress was found~\cite{junPRB2008}.  This negative edge stress was observed to cause wrinkling in free standing graphene nanoribbons~\cite{shenoyPRL2008}. In contrast, we did not observe any compression-induced buckling due to edge stresses in any of our simulations, nor were such compressive stresses found to cause buckling in recent MD simulations of monolayer MoS$_{2}$~\cite{jiangJAP2013} nanoribbons.  One possible reason for this is due to the fact that the bending modulus of MoS$_{2}$ has been reported both theoretically~\cite{jiangNano2013}, and experimentally~\cite{cooperPRB2013,gomezAM2012,bertolazziACSNANO2011} to be about 7-10 times that of monolayer graphene.

\begin{figure} \begin{center} 
\includegraphics[scale=0.6]{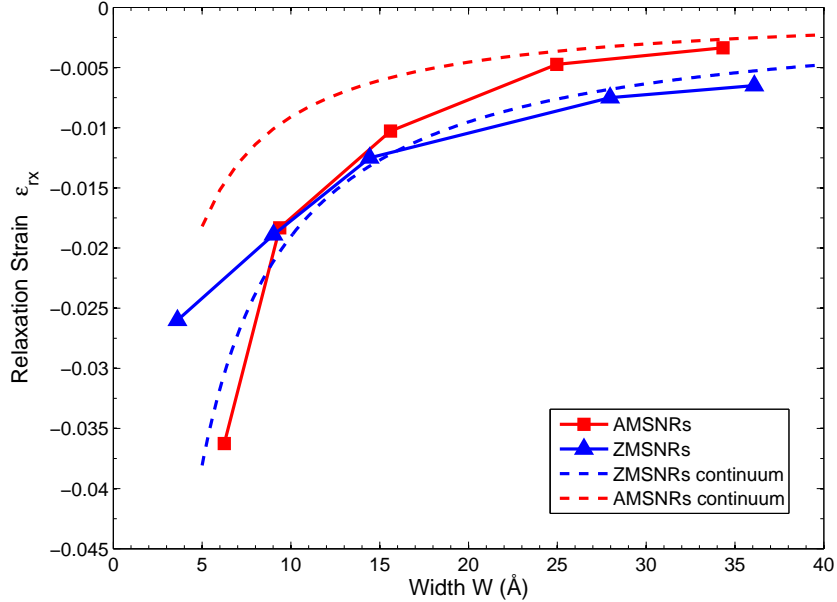}
\caption{(Color online) Stress free (relaxation) strain $\epsilon_{rx}$ as a function of nanoribbon width $W$ for AMSNRs and ZMSNRs as compared to the continuum model presented in Lu and Huang~\cite{luPRB2010}.}
\label{fig4} \end{center} \end{figure}

We plot in Fig. \ref{fig4} the stress free (relaxation) strain $\epsilon_{rx}$ in both AMSNRs and ZMSNRs as a function of nanoribbon width $W$, where the stress-free strain is compressive for both chiralities, which means that in the absence of any externally applied forces or strains, the nanoribbon becomes shorter due to the tensile edge stress observed in Fig. \ref{fig2}.  We also compared our results with the continuum model derived by~\citet{luPRB2010}, who found that the relaxation strain $\epsilon_{rx}$ follows the simple analytic expression
\begin{equation}\label{eq:relax1} \epsilon_{rx}=-\frac{2f}{YW},
\end{equation}
where $f$ is the edge stress, $Y$ is the 2D Young's modulus and $W$ is the nanoribbon width.  We computed the 2D Young's modulus $Y$ of monolayer MoS$_{2}$ by taking the second derivative of the potential energy, and normalizing it by the area $WL$, or the width times the length.  The value was determined to be 8.54~eV/\AA$^{2}$ (or 136.64~N/m) and 8.37~eV/\AA$^{2}$ (or 133.92~N/m) for AMSNR and ZMSNR respectively from our simulations, which are in agreement both with previous experiments~\cite{cooperPRB2013,bertolazziACSNANO2011}, and MD simulations~\cite{jiangJAP2013}. Fig. \ref{fig4} shows that not surprisingly, the continuum model fails to predict $\epsilon_{rx}$ for very small width nanoribbons (i.e. smaller than 1 nm width), while agreeing with the DFT results as the width increases. The relaxation strain in Fig. \ref{fig4} has a crossover for AMSNR and ZMSNR around a width of 1 nm.  The reason for this is because for such small nanoribbon widths, there is no bulk region, i.e. if an individual atom in the Mo plane, or equivalently an individual atom in the S plane is considered, the hexagonal HCP structure is incomplete.  Therefore, the relaxation strain is dominated by the interaction of the two free edges for either armchair or zigzag chiralities.  However, as the nanoribbon widths increase beyond about 20~\AA, it can be observed that the relaxation strain for the AMSNRs is about half that seen in ZMSNRs.  This is physically expected due to the fact that the tensile edge stress for ZMSNRs is about double that of AMSNRs, as shown in Fig. \ref{fig2}. 

\begin{figure} \begin{center} 
\includegraphics[scale=0.8]{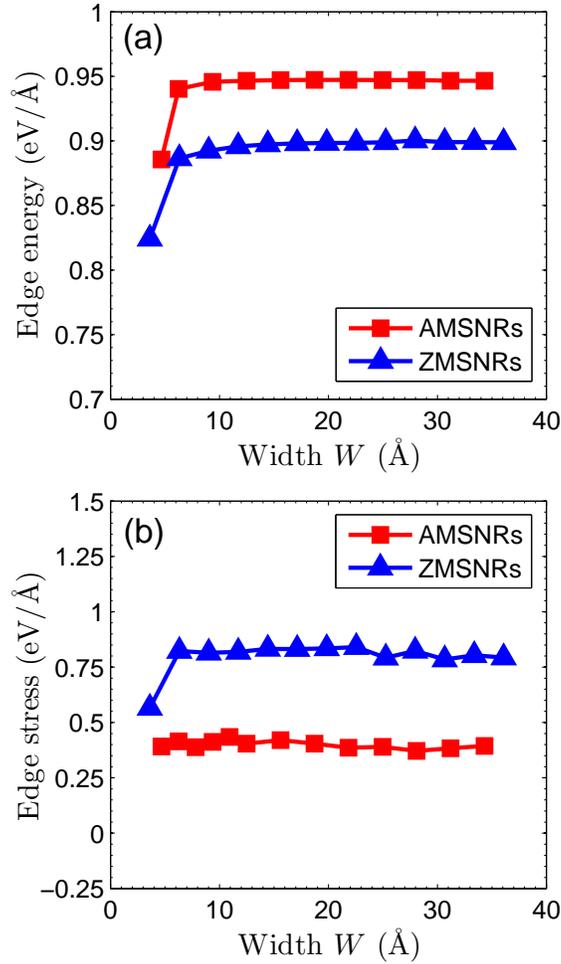}
\caption{(Color online)(a) Edge energy and (b) edge stress as functions of ribbon width $W$ for both AMSNRs and ZMSNRs.}  
\label{fig3} \end{center} \end{figure}

We finally investigate the influence of width on the edge properties.  As shown in Fig. \ref{fig3}(a), once nanoribbons of either chirality become wider than about 1 nm, the edge energy is observed to show very little change as the width increases, and converges to 0.95eV/\AA~for AMSNRs and 0.90eV/\AA~for ZMSNRs as shown in Fig. \ref{fig3}(a).  The edge stress also converges quickly for increasing nanoribbon widths to 0.39eV/\AA~and 0.79eV/\AA~for AMSNRs and ZMSNRs, respectively. Quantitatively, AMSNRs have higher edge energies but lower edge stresses comparing to ZMSNRs. Fig. \ref{fig3} also shows that, similar to that observed in Fig. \ref{fig4}, for ultra narrow MoS$_{2}$ nanoribbons (i.e. one hexagonal ring width) of either armchair or zigzag chirality, the edge energy and edge stress are smaller because the two edges are close to each other and thus edge-edge interactions substantially impact the edge properties.


In summary, we used density functional theory calculations to study the edge energy and edge stress of monolayer armchair and zigzag MoS$_{2}$ nanoribbons.  Both chiralities were found to exhibit a positive edge stress, which implies that their minimum energy configuration is one of compression, where the Mo-S bond lengths are shorter than for bulk monolayer MoS$_{2}$.  The zigzag nanoribbons were found to have a larger edge stress, but a smaller edge energy than the armchair nanoribbons, and for both chiralities the edge energy and stress were found to essentially converge once the nanoribbons become wider than about 1 nm.

ZQ and HSP acknowledge the support of the Mechanical Engineering Department at Boston University.  PC and HSP acknowledge the support of the NSF through grant CMMI-1234183.  All authors acknowledge the assistance of Dr. Jin-Wu Jiang with SIESTA.

\bibliography{biball}

\end{document}